# Thermal transport at a nanoparticle-water interface: A molecular dynamics and continuum modeling study

Ali Rajabpour[a,*], Roham Seif[a,b], Saeed Arabha[a], Mohammad Mahdi Heyhat[c], Samy Merabia[d] and Ali Hassanali[e,*]

[a] Advanced Simulation and Computing Laboratory, Imam Khomeini International University, Qazvin, Iran.

[b] Dipartimento di Energia, Politecnico di Milano, Via Lambruschini 4, 20156, Milano, Italy

[c] Faculty of Mechanical Engineering, Tarbiat Modares University, Tehran, 1411713116, Iran.

[d] Institut Lumière Matière UMR 5306 CNRS Université Claude Bernard Lyon 1, Lyon, France.

[e] International Centre for Theoretical Physics (ICTP), Trieste, Italy.

*E-mail: rajabpour@eng.ikiu.ac.ir, ahassana@ictp.it

**Abstract**

Heat transfer between a silver nanoparticle and surrounding water has been studied using molecular dynamics (MD) simulations. The thermal conductance (Kapitza conductance) at the interface between a nanoparticle and surrounding water has been calculated using four different approaches: transient with/without temperature gradient (internal thermal resistance) in the nanoparticle, steady-state non-equilibrium and finally equilibrium simulations. The results of steady-state non-equilibrium and equilibrium are in agreement but differ from the transient approach results. MD simulations results also reveal that in the quenching process of a hot silver nanoparticle, heat dissipates into the solvent over a length-scale of ~ 2nm and over a timescale of less than 5ps. By introducing a continuum solid-like model and considering a heat conduction mechanism in water, it is observed that the results of the temperature distribution for water shells around the nanoparticle agree well with MD results. It is also found that the local water thermal conductivity around the nanoparticle is greater by about 50 percent than that of bulk water. These results have important implications for understanding heat transfer mechanisms in nanofluids systems and also for cancer photothermal therapy, wherein an accurate local description of heat transfer in an aqueous environment is crucial.

# 1.Introduction

Today, investigating heat transfer in the vicinity of colloidal nanoparticles, i.e nanoparticles surrounded by water, is of great importance for various applications [1-4]. These applications include on one hand nanoparticle based cancer treatments, and on the other hand, the thermal transport properties of nanofluids, which are suspensions of metallic nanoparticles in water and are known to display enhanced thermal conductivity [5-8].

Additionally, in biological systems, the required time and the quenching condition of heated proteins is essential in order to keep their proper function, and this raises some interesting questions regarding the measurement and calculation of interfacial heat transfer between protein and water[1][2][3]. There is also a growing interest in understanding the interactions between inorganic interfaces and biological systems and heat transfer within this context becomes an important factor[4]–[6].

In the studies of interfacial heat transfer between a nanoparticle as a solid and water as a fluid, a discontinuity (jump) between the temperatures of two bodies has been observed. This temperature discontinuity plays a significant role in thermal resistance between the particle and water at the nanoscale. The temperature jump has been observed by Kapitza at first in 1941 in the study of heat transfer between a metal and liquid Helium[7]. The reason of the existence of a temperature jump is related to the difference in acoustic impedance (sound velocity × density) at two sides of the interface [8]. Many studies on thermal resistance have been performed involving both experiments as well as computational simulations at the atomic-scale. Although performing experiments at these small length scales is challenging, some worthwhile works have been recently conducted such as the measurement of Kapitza resistance between silicon and superfluid helium by Ramiere *et al.*[9]. In another experiment, Zhenbin *et al.* utilized the thermo-reflectance method to measure the thermal resistance in both hydrophobic and hydrophilic interfaces and estimated the thickness of a possible vapor layer around the hydrophobic nanoparticle [10].

One of the well-known standard methods for investigating the Kapitza resistance or the inverse of that, the interfacial thermal conductance, are molecular dynamics simulations. Note that there are four main different MD simulations techniques that allow us to calculating the interfacial thermal conductance (i), (ii): Transient non-equilibrium molecular dynamics (TNEMD) methods for obtaining interfacial thermal conductance with/without considering a temperature gradient (internal thermal resistance) in a nanoparticle, (iii): Steady-state non-equilibrium method (SNEMD) and finally, (iv): Equilibrium (EMD) method. Regarding the advantages and disadvantages of each of these methods, TNEMD methods are inspired by experimental measurements of interfacial thermal conductance. But as in these latter methods, the nanoparticle temperature changes with time, and hence determining an exact value of temperature for which the interfacial thermal conductance is measured, is quite challenging. In the SNEMD method, nonlinear temperature effects from hot and cold thermostat regions can cause artificial effects in the simulation. On the other hand, in the EMD method an equilibrium temperature is considered for the whole system thus removing any spurious effect related to the use of two thermostats.

There have been several theoretical studies aimed at studying heat transfer at interfaces, and we briefly review some of these works. Lervik *et al.* studied interfacial thermal conductance between an alkane nanodroplet and water based on TNEMD with/without considering internal thermal resistance in the nanoparticle. They revealed that the interfacial thermal conductance increases with decreased particle diameter[11]. The TNEMD method has been used to study heat transport between a protein and water and thus permitting in elucidating the key role of the interfacial thermal conductance in the thermal relaxation of biomolecules [12]. Barrat and Chiaruttini studied the interfacial thermal conductance between a solid and liquid and investigated the influence of surface wetting by employing the SNEMD and EMD methods[13]. The functional dependence of the Kapitza resistance on the strength of solid-liquid interactions have been rationalized as exponential and power law forms, respectively for non-wetting and wetting liquids[14]. Based on a work by Tascini et al. an investigation on the influence of the wetting characteristics and the nanoparticle size on the interfacial thermal transport between a nanoparticle and fluid, it was found that optimum interfacial heat transport can be realized by combining strong interfacial interactions and large interfacial curvatures. Xu *et al.* studied the role of liquid layering on the liquid–solid interface on the interfacial thermal resistance. Their findings showed that a layering phenomenon is not an effective factor in the thermal transport of nanofluids [15]. The heat transfer mechanism of a gold nanoparticle suspended in water or in octane has been investigated by Merabia *et al.* employing SNEMD method with very high heat fluxes and the interfacial thermal resistance between gold nanoparticle and water/octane has been analyzed[16]. Investigating the surrounding fluid for two cases of a nanoparticle and a planar wall utilizing SNEMD method, Merabia *et al.* showed that the water around the heated nanoparticle can be enhanced above boiling point without any phase change. On the other hand, when the planar wall was heated, the surrounding water converted into vapor and this vapor layer causes a drop in the heat flux value[17]. Wang *et al.* using SNEMD approach examined the influence of nanoscale roughness and wetting on Kapitza resistance at the liquid-solid interface. They showed that the most effective factor on Kapitza resistance is the strength of liquid- solid atomic bonding. As expected, they illustrated that Kapitza resistance decreases with roughness and wetting [18].

Hu and Sun by performing SNEMD simulations, investigated the effect of nanopatterning the interface on the value of the Kapitza resistance between water and gold during boiling. They observed that water boiling has no great effect on the interfacial Kapitza resistance[19]. Pham *et al.* studied the role of pressure on Kapitza resistance of water/gold and water/silicon interfaces via SNEMD method. They found that pressure plays a key role in interfacial thermal conductance at hydrophobic solid surfaces[20]. Vo and Kim calculated the dependence of Kapitza resistance between a variety of metallic surfaces and water with SNEMD approach. They revealed that using a realistic description of liquid water instead of a simple Lennard-Jones (LJ) liquid, noticeable effects on the results of interfacial thermal resistance were found[21]. The effects of wetting and nanoparticle diameter on the thermal transport considering temperature gradient in the particle have been analyzed utilizing SNEMD by Tascini *et al.* It was found that due to very large curvature displayed by the studied nanoparticles, this factor is important in nanoparticle and fluid interactions[22]. The influence of Kapitza resistance on the droplet evaporation rate from a heated surface using a multiscale combination of SNEMD and continuum theory has been investigated by Han *et al.* [23], [24].

Wu *et al.* using EMD approach and ultrafast transient absorption experiments observed the mechanism of thermal transport across surfactant layers of gold nano-rods in aqueous solution [31]. Their MD results showed that the interface between surfactant layers and gold nano-rods noticeably hinders heat flow [25]. Moreover, Soussi *et al.* performed EMD simulations to investigate the role of polymer coating on the gold-water interfacial thermal conductance. According to their results, the coating polymer caused an increase in the interfacial thermal conductance[26].

With respect to the above mentioned studies, all of the four MD approaches have been widely used for calculating the interfacial thermal conductance in different physical situations. So far, and despite their wide use, no comparison has been made among all these methods in a specific case in order to compare and contrast the predictions of each method. The purpose of this article is threefold: First, in this work, we compare TNEMD, SNEMD and EMD methods for calculating the Kapitza resistance at the interface of a silver nanoparticle and water. Second, in the aim of investigating the temperature of different water shells around the nanoparticle when it cools down, the length and timescales associated with heat dissipation are determined. Finally, in an attempt to explore the relevant heat transport mechanisms, a finite element method is used to solve the heat conduction equation for the nanoparticle and its surrounding water and the results are compared to molecular dynamics results.

## 2. Molecular Dynamics Details

Our model system consists of a silver nanoparticle immersed in water. Water has been modeled using the empirical potential TIP4P/2005 [27]. The silver nanoparticles have a spherical shape with a diameter of 1.7nm (225 Ag atoms). Water molecules have been first placed around the nanoparticle and randomly distributed around it. In this study, a simulation box including 6254 water molecules with dimensions of 6nm×6nm×6nm is considered. Periodic boundary conditions are applied in all three spatial directions. The nanoparticle was constrained in the center of the simulation box using a spring with a constant of $100\ kcal/(mol.\text{Å}^2)$. Interatomic forces in silver nanoparticle are modeled using the Lennard-Jones (LJ) potential parametrization taken from Ref. [28]. Water-Ag interactions are calculated according to combination rules of Lorentz-Berthelot. Table 1 presents the LJ parameters, i.e., ε and σ for different atomic interactions.

Table 1. LJ parameters for non-bonding interactions[28]

| Interaction Type | ε (kcal/mol) | σ(Å) |
|---|---|---|
| Ag-Ag | 7.95 | 2.64 |
| $H_2O$ - $H_2O$ | 0.182 | 3.16 |
| Ag - $H_2O$ | 1.21 | 2.9 |

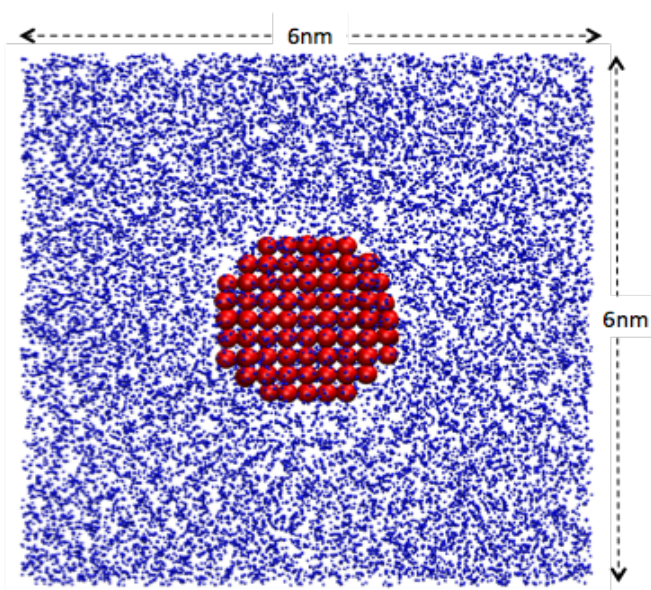

Figure 1: Atomistic structure of a silver nanoparticle surrounded by water. The nanoparticle diameter is about 1.7nm and the box length is about 6nm with periodic boundary conditions in all three dimensions.

The geometry of a silver nanoparticle surrounded by the base fluid of water is illustrated in Figure 1. Molecular dynamics simulation were performed first in the NPT ensemble in order to reach to desired pressure (i.e. 1 bar) and temperature (i.e. 300 K) using a time step of 1 fs. Then the system was run in NVT ensemble for 0.1 ns. After reaching equilibrium, interfacial thermal conductance is obtained using different approaches as alluded to earlier such as TNEMD, SNEMD and EMD, which will be discussed in detail in the following sections. All MD simulations were performed using the open-source Large-scale Atomic-molecular Massively Parallel Simulator (LAMMPS) Package [29].

**3. Results**

*3.1. TNEMD simulations*

In this approach, the temperature of the nanoparticle is initially maintained at 400 K, while the water temperature is maintained at 300 K by using Nose-Hoover thermostats for 50 ps. Then, by switching off the nanoparticle thermostat while maintaining the water thermostat, the equilibration of the nanoparticle temperature is recorded over the course of the simulation. Considering the small size of the silver nanoparticle and the large statistical fluctuations of the temperature, the MD results are averaged over 400 independent trajectories each corresponding to a simulation time of 25 ps. This simulation set up is analogous in spirit to experiments that use laser pulses for heating of a nanoparticle[30]. In order to calculate the interfacial thermal conductance in this approach, we will use two heat transfer models which will be discussed in the following sections.

*3.1.1. Negligible internal resistance in nanoparticle (Lumped model)*

In this model, the temperature distribution within the nanoparticle is assumed to be uniform during the cooling process, this model being known as the lumped capacitance model. Hence, the energy balance equation is written for the nanoparticle as follows:

$$mc_p \frac{dT_{np}}{dt} = -AG[T_{np} - T_W]; T_{np}(t=0) = T_i \quad (6)$$

where $T_{np}$, $c_p$, m and A are the temperature, specific heat capacity, mass and surface area of the nanoparticle, respectively. G denotes the interfacial thermal conductance between surrounding water and the nanoparticle. $T_i$ is the initial temperature of the nanoparticle and $T_W$ is considered to be the surrounding water temperature which is assumed to be constant as a consequence of the use of the applied thermostat. By solving Eq.6, we have:

$$\frac{T_{np}(t) - T_w}{T_i - T_w} = exp\left(\frac{-t}{\tau}\right); \tau = mc_p/(AG) \quad (7)$$

where $\tau$ is the nanoparticle relaxation time. Figure 2 presents the simulation results of nanoparticle and water temperatures during the transient cooling process. In order to determine whether there is a significant temperature gradient or an internal thermal resistance in the nanoparticle or not, the nanoparticle is divided in two parts: a core and a shell. The core and the shell are here defined according to the positions of the atoms: 0<r<0.6 nm and r >0.6 nm, respectively. It is evident from Figure 2 that the shell temperature and the whole particle temperature essentially coincide. Hence, it is clear that there is no significant internal thermal resistance within the nanoparticle. Based on fitting the simulation results to the exponential function in Eq.7, G is obtained as 3.55×10$^8$ W/m$^2$K and $\tau$ is computed to be 2.31 ps.

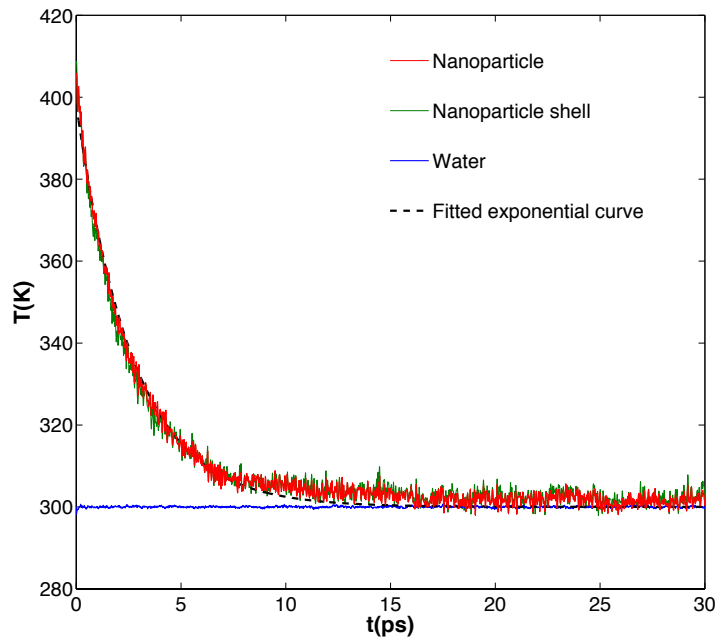

Figure 2: Transient thermal relaxation of hot nanoparticle surrounded by water. The initial temperature of the nanoparticle was set to 400 K and the water temperature was kept at 300 K during relaxation. The temperatures were measured for the whole nanoparticle (0 < r ≤ R) and also for a nanoparticle shell (0.6 nm ≤ r ≤ R) during relaxation. The black dash-line is an exponential curve fitted to nanoparticle temperature.

### 3.1.2. Finite internal resistance in the nanoparticle

According to the lumped model in the previous section, it has been shown that the temperature gradient or the internal thermal resistance in the nanoparticle is small enough to be neglected. In order to reinforce this observation, we consider another model. In some studies [12] [11], a model has been used that considers the temperature gradient in the nanoparticle especially in the outer shell region. In this model, to calculate the interfacial thermal conductance between water and the nanoparticle, the equation of heat diffusion is defined as follows[11]:

$$\frac{\partial^2 (rT_{np})}{\partial r^2} = \frac{1}{\alpha_{np}} \frac{\partial (rT_{np})}{\partial t} \tag{8}$$

$$T_{np}(r, t = 0) = T_i \tag{9}$$

$$-k_{np} \frac{\partial T_{np}}{\partial r}\Big|_{r=R} = G(T_{np}(R, t) - T_W) \tag{10}$$

$$\frac{\partial T_{np}(r, t)}{\partial r}\Big|_{r=0} = 0 \tag{11}$$

where $\alpha_{np}$ is the nanoparticle thermal diffusivity that is equal to $\frac{k_{np}}{\rho c_p}$. The boundary condition in Eq.10, shows the equality of heat conduction in the nanoparticle and interfacial thermal transport at the surface of nanoparticle which has a critical role in this model. From the analytical solution of equations 8 to 11, the nanoparticle temperature can be obtained as follows [11], [31]:

$$\frac{T_{np}(r,t) - T_W}{T_i - T_W} = 4 \sum_{n=1}^{\infty} \frac{\sin(\lambda_n) - \lambda_n \cos(\lambda_n)}{2\lambda_n - \sin(2\lambda_n)} \exp\left(\frac{-\lambda_n^2 t}{\tau}\right) \frac{R}{\lambda_n r} \sin\left(\frac{\lambda_n}{R} r\right) \tag{12}$$

$$1 - \lambda_n \cot(\lambda_n) = \frac{GR}{k} = Bi \quad ; n=1,2,3,... \tag{13}$$

$$\tau = \frac{R^2}{\alpha} \tag{14}$$

The Bi number in Eq.13 is a dimensionless parameter that indicates the ratio of transferred heat between water and the nanoparticle and heat conduction within the nanoparticle. By fitting MD simulations results to Eq.12, the value of the interfacial thermal conductance is calculated as 5.74×10⁸ W/m²K. In addition, the Bi number and the silver nanoparticle thermal conductivity are found to be 2.68 and 0.18 W/mK, respectively. The resulting value of thermal conductivity for the silver nanoparticle is low in comparison to what has been recently reported for phonon thermal conductivity of bulk silver (4 W/mK [32]) and these results suggest that the finite thermal resistance model is not suitable for analyzing the cooling process of silver nanoparticle. As a consequence of small nanoparticle dimension as compared to the long phonon free path in silver (2 nm to 62 nm [32]), heat conduction within the nanoparticle is likely in the ballistic regime, so assuming a diffusive heat transfer inside the nanoparticle and therefore considering a local equilibrium for defining local temperature gradients is certainly not appropriate.

*3.2. SNEMD simulations*

In this approach, the nanoparticle and the water molecules around the simulation box boundaries are steadily maintained at a temperature of 395 K using a hot thermostat and 300 K using a cold thermostat, respectively, as shown in Figure 3a. The required heat fluxes for maintaining these two heat baths at those steady-state temperature values are shown in Figure 3b. Then, by defining concentric shells having a thickness of 0.2nm around the nanoparticle, the temperature profile in water may be resolved as shown in Figure 3a. In order to calculate the interfacial thermal conductance between the nanoparticle and the first water shell around it the following equation is used:

$$G = \frac{|dE/dt|}{A\Delta T} \tag{15}$$

where, dE/dt, A and $\Delta T$ are the heat flux, surface area of the nanoparticle and the temperature drop between the nanoparticle and the closest water shell around it, respectively. Figure 3b. shows the accumulative energy added to the nanoparticle and subtracted from the water versus simulation time. The equality between the slopes of these energy diagrams shows the conservation of energy in our system. It is noticeable that the temperature profile in Fig 3a was averaged over one million time steps (1 ns) in the course of four different MD runs (four million time steps in total). The interfacial thermal conductance based on this model is obtained as G=4.24×10$^8$ W/m$^2$K.

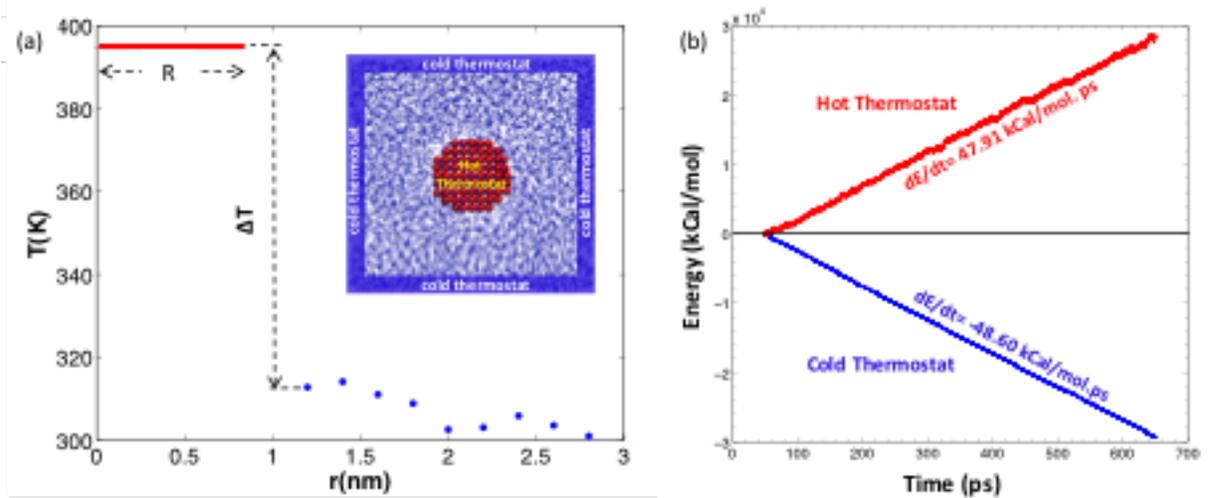

Figure 3: SNEMD simulation setup: (a) Temperature profile between hot thermostat (nanoparticle) and cold thermostat (water molecules at the edges of the simulation box). (b) Accumulative energy added to the nanoparticle and subtracted from water. The slopes of the accumulated energy as a function of time are considered as the stationary heat flux.

*3.3. EMD simulations*

In this approach, the interfacial thermal conductance is obtained based on fluctuations of the temperature difference between the nanoparticle and surrounding water (at cut-off distance of the nanoparticle and water atomic interactions). Rajabpour and Volz have shown that the interfacial thermal conductance (or the inverse of that, Kapitza resistance $R_K$) in the linear response regime can be calculated from the following equation[33]–[35]:

$$G^{-1} = R_K = \frac{A}{k_B T_0^2} \int_0^\infty \langle \Delta T(t) \Delta T(0) \rangle dt \qquad (16)$$

where $k_B$ is the Boltzmann constant, $\langle \ \rangle$ denotes the phase ensemble average and $\Delta T(t)$ is the instantaneous temperature difference between the nanoparticle and the first water shell around the nanoparticle with thickness of 1 nm (corresponding to the cut-off distance of LJ potential for silver-water atomic interactions). *A* is the surface area of the nanoparticle. The temperatures of the nanoparticle and that of water are recorded during one million time steps in order to calculate the auto-correlation function $\langle \Delta T(t) \Delta T(0) \rangle$. Figure 4 shows an example of auto-correlation function of the temperature difference between the nanoparticle and water calculated at equilibrium, and that which is averaged over four different MD runs. It is evident from Figure 4, that this function has decayed to zero in approximately under 10 ps. The integral of this function gives the interfacial thermal resistance value (green curve).

The resulting interfacial thermal conductance based on this approach is equal to G = 4.6×10$^8$ W/m$^2$K. In this model, it is not necessary to apply a temperature difference to the system in comparison to other approaches which is an advantage of EMD method as the interfacial thermal conductance value can be determined from equilibrium fluctuations. In addition, there is no need to calculate the complicated expression of the microscopic heat flux formula in this approach, considering the fact that in Eq.16, only the computation of the temperature difference between the two bodies across the interface is required.

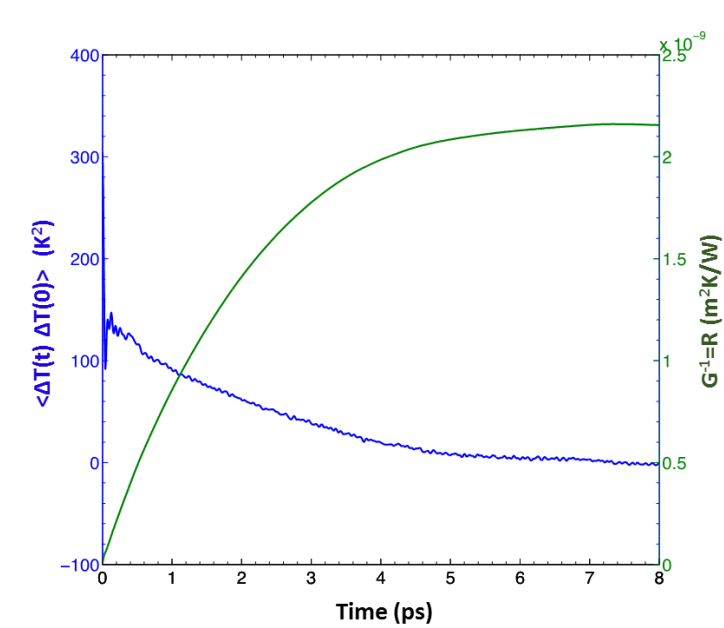

Figure 4: Equilibrium MD simulation results wherein left Y-axis: Temperature difference auto-correlation function versus time. Right Y-axis: Inverse of the interfacial thermal conductance (i.e. interfacial thermal resistance), which is proportional to the integral of temperature difference auto-correlation function.

### 3.4. Comparing the MD approaches

The results of the different MD techniques to calculate the interfacial thermal conductance with the corresponding statistical uncertainties are summarized in Table 2. It is clear that for all the four different methods employed, the values of the interfacial thermal conductance predictions are in the same order of magnitude. However, the results of SNEMD and EMD methods are in better agreement with each other (their difference being less than 10%), compared to what is obtained with the transient methods. More specifically, TNEMD with the lumped approximation technique predicts a value of the conductance 25% lower than that calculated in EMD while TNEMD with finite internal resistance predicts a value of the conductance 25% higher than that of the EMD approaches. The reason for the discrepancy between EMD and TNEMD results may be related to the methodology of TNEMD techniques wherein the nanoparticle is firstly heated and then allowed to cool down, and therefore determining an exact value of the temperature for interfacial thermal conductance measured is not possible. The difference between the results predicted by the two TNEMD techniques (lumped and finite internal resistance) was also previously reported by Lervik *et al.* for the interfacial thermal conductance between an alkane nanodroplet and water [11].

Table 2: Interfacial thermal conductance G, calculated from the four different MD methods discussed in the text

| Method | $G$ (W/m$^2$K) $\times 10^8$ |
|---|---|
| TNEMD (lumped method) | 3.55 ± 0.06 |
| TNEMD (Finite internal resistance) | 5.74 ± 0.09 |
| SNEMD | 4.24 ± 0.32 |
| EMD | 4.6 ± 0.15 |

*3.5. Heat dissipation in surrounding water*

Now that we have discussed and compared the different possible methods to extract the value of the Kapitza resistance, we analyze in detail the local temperature profile in water. In the TNEMD technique presented in section 3.1, the water temperature was maintained at 300 K using a cold thermostat during the cooling process of the heated nanoparticle. The assumption of a uniform temperature profile for the whole surrounding water is questionable. Stoll *et al.*

using an ultra-probe experiment showed the existence of a temperature gradient taking place in the surrounding water shells. In this section, for determining the amount of temperature increase in the surrounding water around the nanoparticle during the cooling process, the water thermostat is switched off and the local temperature field in water shells around the nanoparticle, are recorded. This allows us to understand the mechanism of dissipation of nanoparticle energy into the solvent in order to ascertain the spatial and temporal scales associated with the cooling process.

The initial temperature of the nanoparticle was set to 400K, while the initial water temperature was 300K. This temperature jump of the nanoparticle is typical of what can be reached in thermoreflectance experiments, when the nanoparticle is heated up by a laser working in the vicinity of the plasmon resonance frequency. The surrounding water was divided into concentric shells with a thickness of 0.2nm. The temperature distribution of water shells versus time is shown in Figure 5a. The presented temperatures are averaged over 2000 MD trajectories. It is clear from the Figure that, the temperature of the first water shell experiences an increase of about 5 K and this gradually drops to below 1 K at a distance r ~ 2 nm. The temperature increase for further water shells is on the order of temperature fluctuations and can be thus neglected. These small increases in temperatures of surrounding water shells may be understood based on the high heat capacity of water that acts like a natural temperature damper (small temperature response to the incoming heat). The peak time of the temperature for the different water shells are shown in Figure 5b. It is seen that this timescale linearly shifts from short to longer time with increasing the shell radius.

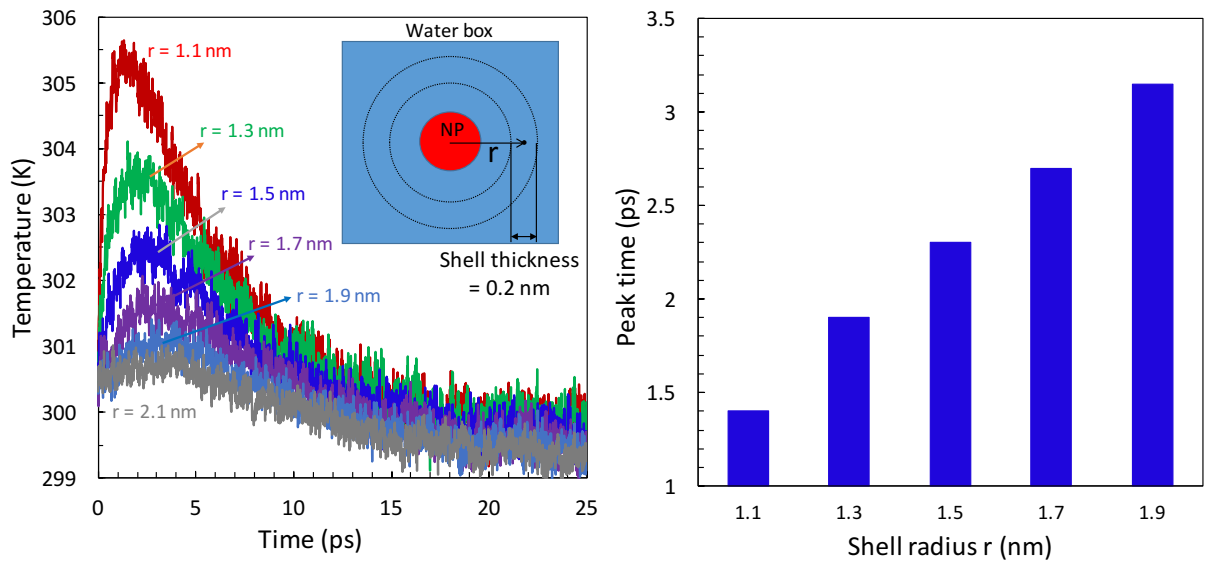

Figure 5: (a) Temperature of the different water shells around the nanoparticles with respect to the simulation time during the thermal relaxation of the nanoparticle. The first shell experiences an increase in the temperature of about 5K, while for shell with radius of about 2nm, there is no clear temperature increase. (b) The corresponding time for the temperature peak in each shell versus the radius of the shell.

*3.6 Heat transport mechanism in surrounding water*

In the previous section, we observed a small increase in water temperature, which persists up to a distance about 2nm away from the nanoparticle. In order to understand this, a simple continuum heat equation model is considered to understand an effective mechanism that describes heat transport within surrounding water well. It is well established from numerous other studies that both hydrophobic and hydrophilic interfaces introduce structuring in the water in close vicinity to them [36]. Indeed, in the case of the silver-water interface, we also see some structuring. The density of water shells directly obtained from molecular dynamics simulation (Figure 6 a and b) clearly shows that the density of the first water shell (r=1.1 nm) is different from the other shells [37] [38]. The thermal behavior of water shells adjacent to the nanoparticle can be described by a *solid-like* model and a thermal conduction mechanism could be considered for describing the heat transfer within those shells [39]. The solid-like model does not imply that there is an ice-like layer of water near the interface. It just implies more structured water near the silver surface. In this context, the transient heat conduction equations for the nanoparticle and surrounding water should be written as coupled equations:

Nanoparticle: (17)

$$\frac{\partial^2 (rT_{np})}{\partial r^2} = \frac{1}{\alpha_{np}} \frac{\partial (rT_{np})}{\partial t} \qquad 0 < r < R$$

$$\frac{\partial T_{np}(r,t)}{\partial r}\bigg|_{r=0} = 0$$

$$\kappa_{np} \frac{\partial T_{np}(r,t)}{\partial r}\bigg|_{r=R} = G(T_{np} - T_W(R,t))$$

$$T_{np}(r,0) = T_i$$

Water around the nanoparticle: (18)

$$\frac{\partial^2 (rT_W)}{\partial r^2} = \frac{1}{\alpha_W} \frac{\partial (rT_W)}{\partial t}$$

$$-\kappa_W \frac{\partial T_W(r,t)}{\partial r}\bigg|_{r=R} = G(T_{np} - T_W(R,t))$$

$$T_W(r \to \infty, t) = T_{W0}$$

$$T_W(R,0) = T_{W0}$$

where $T_{np}$ is the nanoparticle temperature, $\alpha_{np}$ is thermal diffusivity of the nanoparticle and R its radius, $T_W(r,t)$ is surrounding water temperature at radius $r$ and time $t$, $\kappa_W$ is thermal conductivity of water and $\alpha_W$ represents the thermal diffusivity of water equals to $\frac{\kappa_W}{\rho_W c_p}$.

$T_i$ is the initial temperature of the nanoparticle that equals to 400 K. $T_{W0}$ is the temperature of water far from the nanoparticle which is assumed to be 300 K based on MD results which show that water shells at a distance greater than $r$ =2 nm, have temperatures approximately around 300 K. In order to solve the coupled equations of 17 and 18, the thermo-physical properties of the nanoparticle and the surrounding water around it are needed. To this aim, we assume that the nanoparticle shares the characteristics of bulk silver. Therefore, the high thermal conductivity of silver result in Eq.18 resulting like a lumped model. The specific heat capacity and thermal conductivity of bulk water are assumed to be 4.2 kJ/kgK and 0.8 W/mK, respectively, obtained from an MD simulation based on TIP4P/2005[40]. The interfacial thermal conductance between the water and the nanoparticle is set to G = 4.6×10$^8$ W/m$^2$K as obtained from EMD calculations. The density of water shells is directly obtained from molecular dynamics simulation as shown in

Figure 6. It can be seen that the density of the first water shell (r=1.1 nm) is different from the other shells. Thus, the thermal conductivity of this water shell may not equal to the bulk water thermal conductivity, but as an initial estimation, the thermal conductivity of bulk water is assumed for this water shell. The temperatures of solid-like model resulted from numerical solution of equations 17 and 18 through a finite element method (FEM) are plotted in Figure 6c. It is evident that there is a qualitative agreement between temperature distribution of water shells around the nanoparticle from MD simulations and FEM results. This can confirm that the continuum solid-like heat diffusion model captures the important physics associated with the length and time-scale associated with the evolution of the water temperature in close vicinity of the nanoparticle, despite the nanoscale structural changes in water induced by the presence of the nanoparticle.

One of the assumptions of the previous analysis was that the thermal conductivity of water at the interface is the same as that of the bulk. In order to assess the sensitivity of our analysis to the choice of the water thermal conductivity, the difference between MD and FEM results for the temperature of the first water shell in various thermal conductivities is computed as follows:

$$S^2 = \frac{1}{N} \sum_{t=0}^{25\ ps} \left(T_{FEM}(t) - T_{MD}(t)\right)^2 \qquad (19)$$

where N is the number of recorded data in the quenching process. As shown in Figure 7, the minimum of S is obtained using a thermal conductivity of k = 1.4 W/mK for the first water shell. Hence, it can be revealed that the thermal conductivity of first water shell around the nanoparticle is likely higher than thermal conductivity of bulk water by about 50%. Qualitatively, this result is not surprising, as the first water shell due to its enhanced structuring should be a better thermal conductor as bulk liquid water. However, the qualitative features of the dissipation mechanism are also well predicted using the bulk thermal conductivity of water.

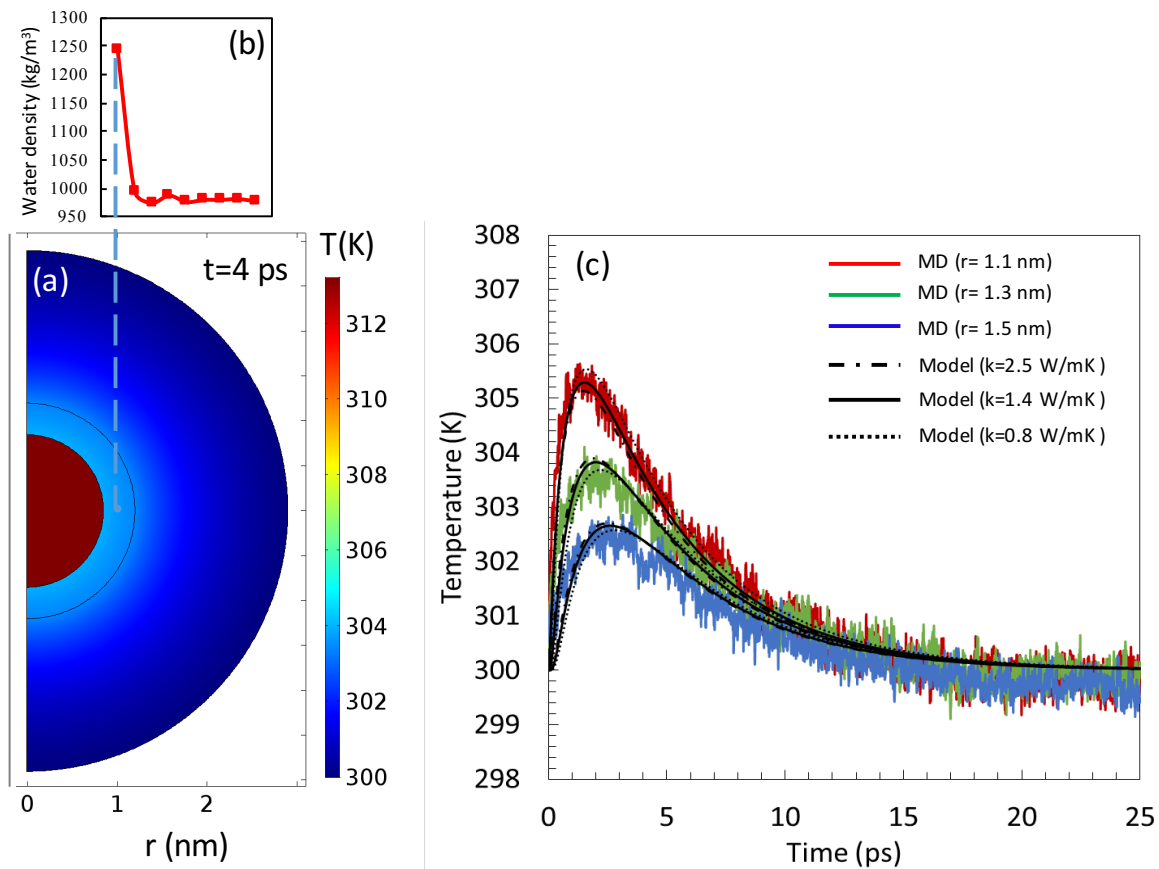

Figure 6: (a): Temperature distribution of water around the nanoparticle at t = 4ps calculated from the continuum model (FEM) (b): Density of surrounding water directly obtained from MD simulation. (c): Temperature of different water shells around the nanoparticle during the thermal relaxation of the nanoparticle. Colored lines are MD results and black dashed lines are continuum model (FEM) calculated with different values of the thermal conductivities for the first water shell.

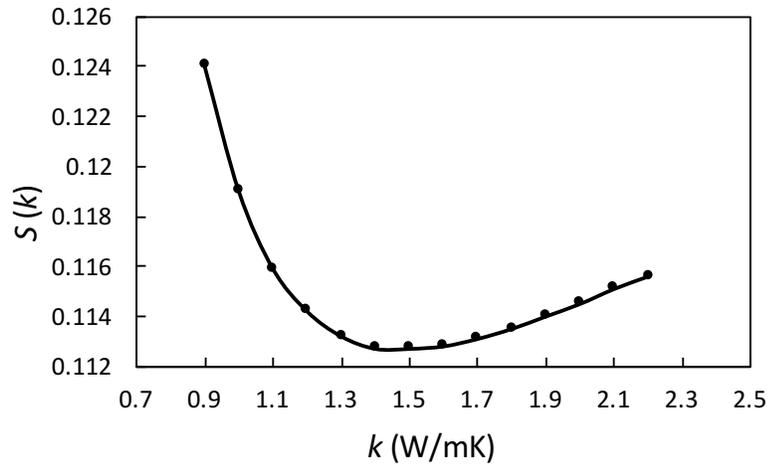

Figure 7: The difference between MD and FEM temperature values (Eq.19) for various thermal conductivities of the first water shell.

**4. Conclusion**

In this paper, the thermal transport mechanisms between a nanoparticle and surrounding water was investigated using molecular dynamics simulations and compared with continuum modeling. At first, the interfacial thermal conductance based on four different MD approaches was calculated. The results from all of four approaches were within the same order of magnitude. SNEMD and EMD approaches showed good agreement (under 10% difference) with each other while transient techniques displayed slightly greater difference (about 25%) with SNEMD and EMD values. MD results also reveal that in the cooling process of a hot silver nanoparticle, the heat dissipates into the solvent over a length-scale of ~ 2nm and over a timescale of less than 5ps.

By introducing a continuum solid-like model and considering a heat conduction mechanism in water, it is observed that the results of the temperature distribution for water shells around the nanoparticle agree with the MD results. Moreover, with introducing a solid-like continuum model and comparing the results with the MD simulations, it is shown that the dominant mechanism in heat transfer in water near the nanoparticle on these timescales involves only conduction and not convection. It is also revealed that the local thermal conductivity of water shell adjacent to the nanoparticle is higher than that of bulk water by about 50%.

Our results have important implications in the understanding and development of engineering devices relevant for heat transport as well as in medical applications for thermal therapy cancer treatment methods. In the future, it will be interesting to examine how the heat dissipation mechanism is sensitive to both the chemical details of the interface being considered as well as the type of solvent.